\documentclass[a4paper,11pt]{article}

\usepackage[T1]{fontenc}
\usepackage[utf8]{inputenc}
\usepackage{lmodern}
\usepackage{graphicx,color}
\usepackage{amsmath,amsfonts}%,amssymb,bbm}
\usepackage{tabularx}

\title{Towards a statistical mechanics of consciousness: maximization of number of connections is associated with conscious awareness}
\author{R. Guevara Erra$^{1}$, D. M. Mateos$^{2}$, R. Wennberg$^{3}$, J.L. Perez Velazquez$^{2}*$ \\
\footnotesize $^1$~{Laboratoire Psychologie de la Perception, CNRS  and Université Paris Descartes, Sorbonne Paris Cité, Paris, France.} \\
\footnotesize $^2$~{Neuroscience and Mental Health Programme, Division of Neurology, Hospital for Sick Children.}\\
\footnotesize{ Institute of Medical Science and Department of Paediatrics, University of Toronto, Toronto, Canada.} \\
\footnotesize$^3$~{ Krembil Neuroscience Centre, Toronto Western Hospital, University of Toronto, Toronto, Canada. }
\\
\tt\footnotesize * jose-luis.perez-velazquez@sickkids.ca
}

\begin{document}

\maketitle
%\tableofcontents

\begin{abstract}
It has been said that complexity lies between order and disorder. In the case of brain activity, and physiology in general, complexity issues are being considered with increased emphasis. We sought to identify features of brain organization that are optimal for sensory processing, and that may guide the emergence of cognition and consciousness, by analysing neurophysiological recordings in conscious and unconscious states. We find a surprisingly simple result: normal wakeful states are characterised by the greatest number of possible configurations of interactions between brain networks, representing highest entropy values. Therefore, the information content is larger in the network associated to conscious states, suggesting that consciousness could be the result of an optimization of information processing. These findings encapsulate three main current theories of cognition, as discussed in the text, and more specifically the conceptualization of consciousness in terms of brain complexity. We hope our study represents the preliminary attempt at finding organising principles of brain function that will help to guide in a more formal sense inquiry into how consciousness arises from the organization of matter

\end{abstract}

\section{Introduction}
How consciousness arises from the organization of matter is a subject of debate that spans several disciplines, from philosophy to physics. Multitude of studies focus on the investigation of patterns of synchrony in brain activity based on magnitudes of a variety of synchrony indices, and thus the search for organising principles of brain function is today more crucial than ever. We sought to identify global features of brain organization that are optimal for sensory processing and that may guide the emergence of conscious awareness. Our results provide a (very simple) answer to the question of what the magnitudes of synchrony indices represent in terms of the structure of brain activity.
Neurophysiological recordings of brain activity demonstrate fluctuating patterns of cellular interactions, variability that allows for a wide range of states, or configurations of connections of distributed networks exchanging information, and support the flexibility needed to process sensory inputs. Recent years have seen a surge in the study of fluctuations in brain coordinated activity, studies that have raised conceptual frameworks such as that of metastable dynamics \cite{kelso1997dynamic} and that have motivated interest in the practical application of assessments of nervous system variability for clinical purposes \cite{garrett2013moment, nenadovic2014phase}. A prominent question is how to describe the organizing principles of this collective activity, which allow features associated with consciousness to emerge. What is the optimal brain organization that allows it to adequately process sensory stimuli and enable the organism to adapt to its environment?  

Previous studies have revealed values of different indicators of brain coordinated activity, such as synchronization, associated with healthy and pathologic states by comparison of baseline values and those in, for instance, unconscious states like coma and epileptic seizures \cite{mormann2000mean, dominguez2005enhanced, shields2007cortical}. These observations prompt the question of 
what physiological organization underlies the specific values of the synchrony indices found in normal alert states and other conditions; in other words, what is special about these values found in conscious states? We believe that we have provided an answer to this question in our work. We propose that there is a certain general organization of brain cell ensembles that will be 
optimal for conscious awareness, that is, for the processing of sensory inputs. As an extension of previous work \cite{velazquez2009finding} where it was proposed that a general organising principle of natural phenomena is the tendency toward maximal – more probable – distribution of energy/matter, we venture that the brain organization optimal for conscious awareness will be a manifestation of the tendency towards a widespread distribution of energy (or, equivalently, maximal information exchange). Whereas we do not deal with energy or information in our work, we instead focus on the number of (micro)states, or combinations of connected signals derived from specific types of neurophysiologic recordings. We use the term “information” in the intuitive sense that normally permeates neuroscience: cell ensembles that are functionally connected to process/exchange information; furthermore, the equivalence between information exchange and energy transactions has been the subject of several 
studies, more specifically in \cite{velazquez2009finding}  (see also \cite{morowitz1955some, smith2008thermodynamics}). The 
question then becomes: how do we capture the nature of these organizations of cell interactions? 

We have followed the classic approach in physics when it comes to understanding collective behaviours of systems composed of a myriad of units: the assessment of the number of possible configurations, or microstates, that the system can adopt.  In our study we focus on the collective level of description and assume that coordinated patterns of brain activity evolve due to interactions of mesoscopic areas ( \cite{ wright1996dynamics, nunez2000toward}). Thus we use several types of brain recordings in conscious and unconscious states, evaluating the number of “connections” between these areas and the associated entropy and complexity. We present evidence that conscious states result from higher entropy and complexity in the number of configurations of pairwise connections. The number of pairwise channel combinations is near the maximum of all possible configurations when the individual is processing sensory inputs in a normal manner (e.g. with open eyes). Our interpretation is that a greater number of configurations of interactions allows the brain to optimally process sensory information, fostering the necessary variability in brain activity needed to integrate and segregate sensorimotor patterns associated with conscious awareness. 
\section{Materials and Methods}

\subsection*{Electrophysiological recordings}

Recordings were analysed from 9 subjects, using magnetoencephalography (MEG), scalp electroencephalography (EEG) or intracranial EEG (iEEG). Three epilepsy patients were studied with MEG; 1 epilepsy patient was studied with iEEG; 3 epilepsy patients were studied with simultaneous iEEG and scalp EEG; and 2 nonepileptic subjects were studied with scalp EEG. For the study of seizures versus alert states, the 3 subjects with MEG recordings and the one with iEEG were used. Details of the patients’ epilepsies and seizure types have been presented in previous studies (MEG patients in \cite{dominguez2005enhanced}; iEEG patients in \cite{velazquez2011experimental}). For the study of sleep versus alert states, the 3 patients with combined iEEG and scalp EEG have been described previously (patients 1, 3, 4 in \cite{wennberg2010intracranial}); the 2 subjects studied with scalp EEG alone had been investigated because of a suspected history of epilepsy, but both were ultimately diagnosed with syncope, with no evidence of epilepsy found during prolonged EEG monitoring. In brief, the MEG seizure recordings were obtained in one patient with primary generalized absence epilepsy, in one patient with symptomatic generalized epilepsy, and in one patient with frontal lobe epilepsy. The iEEG seizure recordings were obtained from a patient with medically refractory temporal lobe epilepsy as part of the patient’s routine clinical pre-surgical investigation. MEG recordings were obtained using a whole head CTF MEG system (Port Coquitlam, BC, Canada) with sensors covering the entire cerebral cortex, whereas iEEG electrodes were positioned in various locations including, in the temporal lobe epilepsy patient, the amygdala and hippocampal structures of both temporal lobes. EEG recordings were obtained using an XLTEK EEG system (Oakville, ON, Canada). The details of the acquisitions varied from patient to patient (e.g., acquisition rate varied from 200 to 625 Hz) and were taken into consideration for the data analyses. The duration of the recordings varied as well: for the seizure study, the MEG sample epochs were of 2 minutes duration each, with total recording times of 30-40 minutes; the iEEG patient sample was of 55 minutes duration. The sleep study data segments were each 2-4 minutes in duration, selected from continuous 24-hour recordings. 
\subsection*{Data analysis}
The only pre-processed data were those of the scalp EEG recordings. These were processed using a Laplacian derivation \cite{nunez2000toward}, to avoid the potential effects of the common reference electrode on synchronization \cite{guevara2005phase}  using the DSC algorithm \cite{kayser2006principal}. Initially a phase synchrony index was calculated from all possible pairwise signal combinations, for which we use the standard procedure of estimating phase differences between two signals from the instantaneous phases extracted using the analytic signal concept via the Hilbert transform. To compute the synchrony index, several central frequencies, as specified in the text and figure legends, were chosen with a bandpass filter of 2 Hz on either side, hence, for one value of the central frequency $f$, the bandpass is $f \pm 2$ Hz. The central frequencies were chosen according to the relevant behavioural states and some analytical limitations– thus for the sleep studies we choose 4 Hz (not lower because the extraction of the instantaneous phase was not optimal for central frequencies lower than 4). To see whether similar results were obtained with different frequencies, we chose others (see figures) provided there was power at those values. The phase synchrony index ($R$) was calculated using a 1-second running window and was obtained from the phase differences using the mean phase coherence statistic which is a measure of phase locking and is defined as $R=\lvert   \langle e^{i \Delta \theta } \rangle   \lvert $ where $ \Delta \theta$ is the phase difference between two signals. This analytical procedure has been described in great detail elsewhere \cite{mormann2000mean, dominguez2005enhanced, guevara2005phase}. The calculation of the index R was done for all possible signal pairs.  The mean value of the $R$ index thus obtained was then estimated for the desired time length. For instance, for the sleep recordings, the time period was the whole episode, which was, as noted above, between 2 and 4 minutes. For the seizure recordings, the periods to obtain the mean synchrony varied depending on the behavioural condition, for instance, in figure \ref{figure1}C the whole ictal event (labelled ‘Sz’) and the initial 10-second portion of it (’10 sec Sz’) were taken for the reason explained in the text and figure legend. The calculation of the number of connected signals and the entropy associated is described in the Results section. We note here that to assess entropy we assume that the different pairwise configurations are equiprobable, thus the entropy is reduced to the logarithm of the number of states, $S = ln C$ (see notation in the main text). However, the estimation of $C$ (the combinations of connections between diverse signals), is not feasible due to the large number of sensors; for example, for 35 sensors, the total possible number of pairwise connections is $[144  2]=10296$, then if we find in the experiment that, say, $2000$ pairs are connected, the computation of $[10296 2000]$ has too large numbers for numerical manipulations, as they cannot be represented as conventional floating point values in, for instance, MATLAB. To overcome this difficulty, we used the well-known Stirling approximation for large $n: ln(n!) = n~ln(n) – n$. The Stirling approximation is frequently used in statistical mechanics to simplify entropy-related computations. Using this approximation, and after some basic algebra, the equation for entropy reads, $S = N~ln (N/N-p) - p~ln (p/ N-p)$, where $N$ is the total number of possible pairs of channels and p the number of connected pairs of signals in each experiment (see Results for details and notation). Because this equation is derived from the Shannon entropy, it indicates the information content of the system as well (37).

In addition to entropy, we used another measure of complexity, the Lempel-Ziv (L-Z) complexity, based in 
the Kolmogorov deterministic complexity \cite{kolmogorov1965three}. This complexity measure the amount of 
non-redundant information in a string by estimating the minimal size of the ''vocabulary'' necessary to 
describe the string \cite{lempel1976complexity}. Strings with high L-Z complexity require a large number of 
different patterns (“words”) to be reproduced, while strings with low complexity can be largely compressed 
with a few patterns employed to eliminate redundancy with no loss of information. For this purpose, values 
of the matrix $B$ (defined in Results) were placed in a one-dimensional vector and its L-Z complexity determined. 

\section{Result}

Guided by proposals that consciousness requires medium values of certain features of cell assemblies, e.g. not too high or low synchrony or correlations \cite{tkacik2014thermodynamics, nenadovic2008fluctuations, nenadovic2014phase}, or halfway between order and disorder \cite{tononi1998complexity}, we chose to quantify the number of possible configurations the brain can adopt in different behavioural conditions. Our basic approach consists in the estimation of the number of possible pairwise connections between recorded brain signals. Signals included MEG, iEEG and scalp EEG recordings; details can be found in Methods. We are limited to pairwise combinations of the signals because of the manner in which phase synchrony is computed – as phase differences between two signals – and we use phase synchronization as the means to determine “connectivity” between the two signals. Once the number of “connected” signals is known, we estimate the entropy of those pairwise combinations. The results obtained with recordings acquired during conscious states are compared with those acquired during unconscious states, which included sleep (all stages) and epileptic seizures. To determine connectivity, we use an accepted approach of computing a phase synchronization index (details in Methods). 

It must be noted that, while many studies use the words synchrony and connectivity as synonymous, in reality phase synchrony analysis reveals only a correlation between the phases of the oscillations between two signals, and not a real connectivity which depends on several other factors (this matter has been covered in detail in \cite{velazquez2009correlations, velazquez2011brain}. Nevertheless, due to the unfeasibility of an accurate, realistic estimation of connectivity which would necessitate individual cell recordings from entire cell ensembles as well as structural connectivity details, we use an accepted version under the assumption that the phase relations may represent, at least, some aspect of a functional connectivity. Hence, in order to evaluate interactions (“connections”), we take each sensor/channel as one “unit”, and define a pair of signals as “connected” if the phase synchrony index is larger than a threshold. The threshold is determined for each individual, and is the average synchrony index in the most normal alert state, the ‘awake eyes-open’ condition, when the individual is fully alert and processing the sensorium in a regular fashion. Because our data include the three recording methodologies aforementioned, we have the opportunity to assess the reproducibility of the results in various types of recordings. While we work at the signal level we will make the reasonable assumption that the MEG and scalp EEG sensors record cortical activity underlying those sensors and thus throughout the text we will use the terms brain signals or brain areas/networks as synonymous. The iEEG, obviously, records signals at the source level. Note that we are not interested in the specific pattern of connectivity among brain sources/areas, but rather in the global states. These points are further discussed in the Discussion section.   
  
Phase synchronization for specific frequencies (details in Methods) is calculated for each pair of channels and a “connectivity” matrix $S$ is obtained, whose entries are the average values of the synchrony index during a certain time period for each pairwise configuration. From this matrix, a Boolean connectivity matrix $B$ is calculated, with 0 entry if the corresponding synchrony index is lower than a threshold, and 1 if higher. We define two channels as “connected” if the corresponding entry in matrix $B$ is 1. Then we use the combinations of connected channels as a ‘complexity’ measure. The total number of possible pairs of channels given a specific channel montage is given by the binomial coefficient   $N = Nc!/2!~(Nc-2)!$  where $Nc$ is the total number of channels in the recording montage, normally $144-146$ in the case of MEG sensors, and between 19 and 35 with iEEG and scalp EEG. For instance, in our MEG recordings we have$ Nc=144$, thus $N=10,296 $possible pairs of connected sensors are obtained. For each subject we calculate p, the number of connected pairs of signals in the different behavioural states, using the aforementioned threshold of the synchrony index (which varies for each subject), and estimate $ C$, the number of possible combinations of those p pairs, using the binomial coefficient again:  $C = N!/p!(N-p)!$  where $N$ is the aforesaid value. In sum, all these calculations represent the relatively simple combinatorial problem we are trying to solve: given a maximum total of $N$ pairs of connected signals, in how many ways can our experimental observation of p connected pairs (that is, the number of 1’s in matrix $B$) be arranged. We then compute the entropy and Lempel-Ziv complexity associated with those p values.
  
\begin{figure}[]  
  \begin{center}
    \includegraphics[scale=1.1]{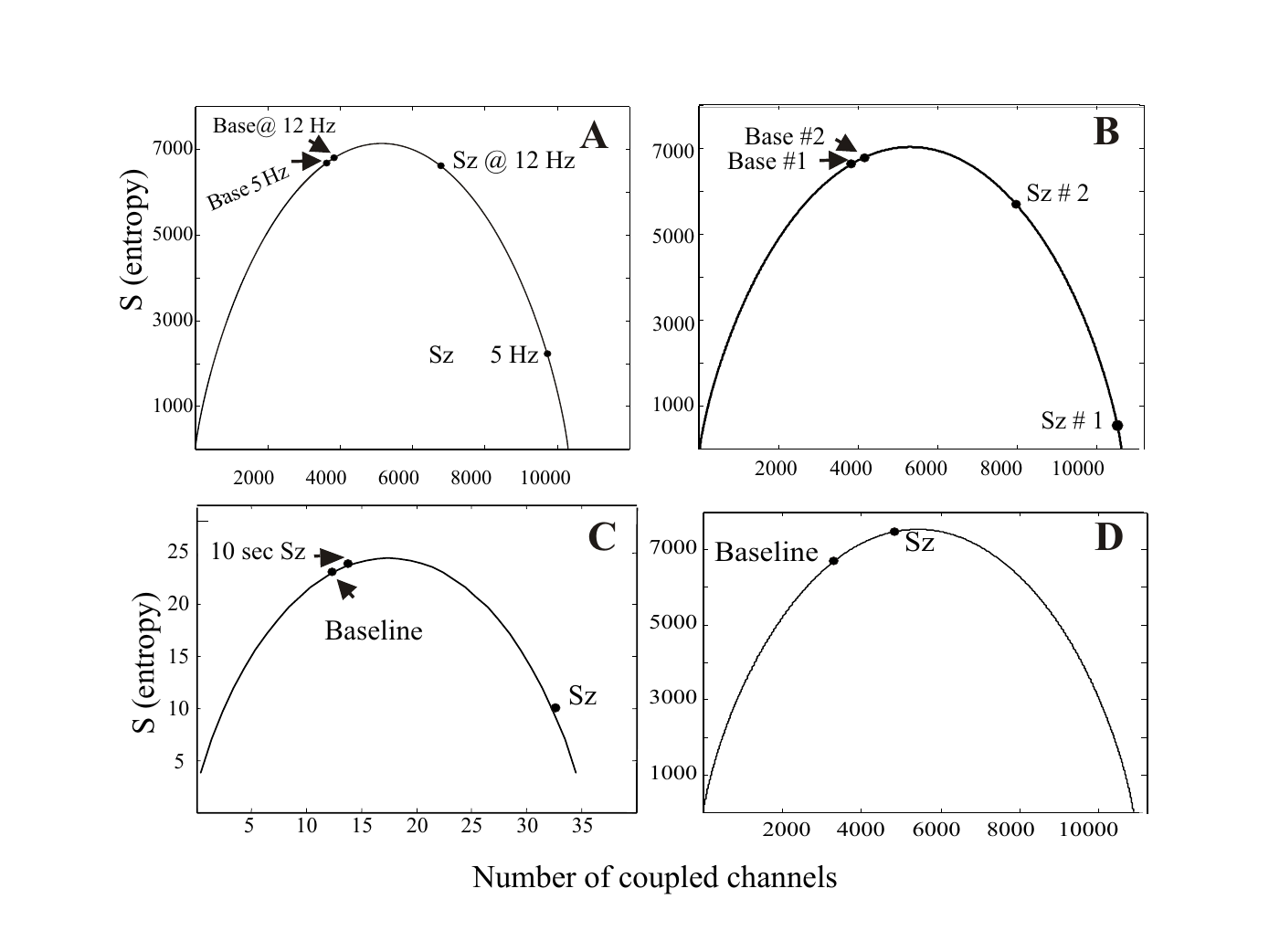}
    
    \caption{Graphs representing the entropy of the number of pairwise configurations of signals in epileptic patients during conscious (baseline) and unconscious (generalised seizure) states. A, derived from MEG recordings in a patient with primary generalised epilepsy, shows entropy associated with a normal alert period (baseline, ‘Base’) and a generalised absence seizure (‘Sz’), estimated from synchrony values at two central frequencies (defined in Methods) of 5 and 12 Hz. The curve in this and other graphs here and in Figure 2 represents the possible entropy values of all possible numbers of pairwise combinations, yielding an inverted U or, for very large numbers, a Gaussian. Note that here as well as in all generalised seizures analysed, the entropy values associated with alert, baseline conditions were closer to the maximum (top of the curve) than those associated with the seizures. B, entropy values of two seizures and their corresponding baseline (‘Base’) activity (computed using a time period of 30-40 minutes before the ictus) in a patient with secondary (symptomatic) generalised epilepsy (MEG recordings). C, derived from iEEG recordings in a patient with temporal lobe epilepsy, shows the entropy during the alert state (‘baseline’), during the initial 10 seconds of the seizure when the patient was still responsive and alert (‘10 sec Sz’), and during the rest of the seizure when it became generalised and the patient was unresponsive (‘Sz’). Note that when the ictus has not yet generalised, the entropy is similar to that of normal alertness. D, another example of a non-generalised seizure in a patient with frontal lobe epilepsy (MEG recordings). }  
    \label{figure1}
  \end{center}  
\end{figure}

Figure \ref{figure1} depicts the entropy ($S$), which, assuming equiprobable states, is the logarithm of the number of states, $S = ln~C$ (see Methods for the estimation of entropy using very large $C$ values) in three epileptic patients. We note that this equation using the natural logarithm allows for a calculation of both the Gibbs and the Shannon entropy, if needed, which differ from $S$ by a constant multiplicative factor k (the Boltzmann constant) and $1/(ln 2)$ respectively. In reality the entropy estimation does not provide any further information, as the main, crucial result is the number of configurations, $C$.  However, we have done it since it is a standard manner to quantify the “complexity” of the number of microstates. The entropy data points are graphed on the curve that represents the entropy of all points in the binomial distribution, where the maximum number of configurations (that is, maximum entropy) occurs in the middle of the graph. Note in that figure that during conscious states, when patients are not having generalised seizures with loss of awareness, the entropy is close to the maximum, whereas entropy is lower (more distant from the top) for the seizure states. The values during the seizures fall on the right-hand side of the graph because, due to the higher synchrony during ictal (seizure) events such that the number of coupled channels (note that the x-axis is the number of connected signals) is larger than during interictal (between seizure) activity, there are fewer pairwise configurations and thus lower entropy. This phenomenon seems associated with the level of consciousness since when the seizures are not generalised (Figure \ref{figure1}C and \ref{figure1}D), and the patients remain responsive and conscious, the entropy values are similar to those of interictal (baseline) activity. Where in the curve the data points are located depends of course on the synchrony index. Because seizures had higher synchrony than interictal periods (“baseline”), the number of coupled signals is greater and the number of combinations is lower. 
  
\begin{figure}[]  
  \begin{center}
    \includegraphics[scale=1]{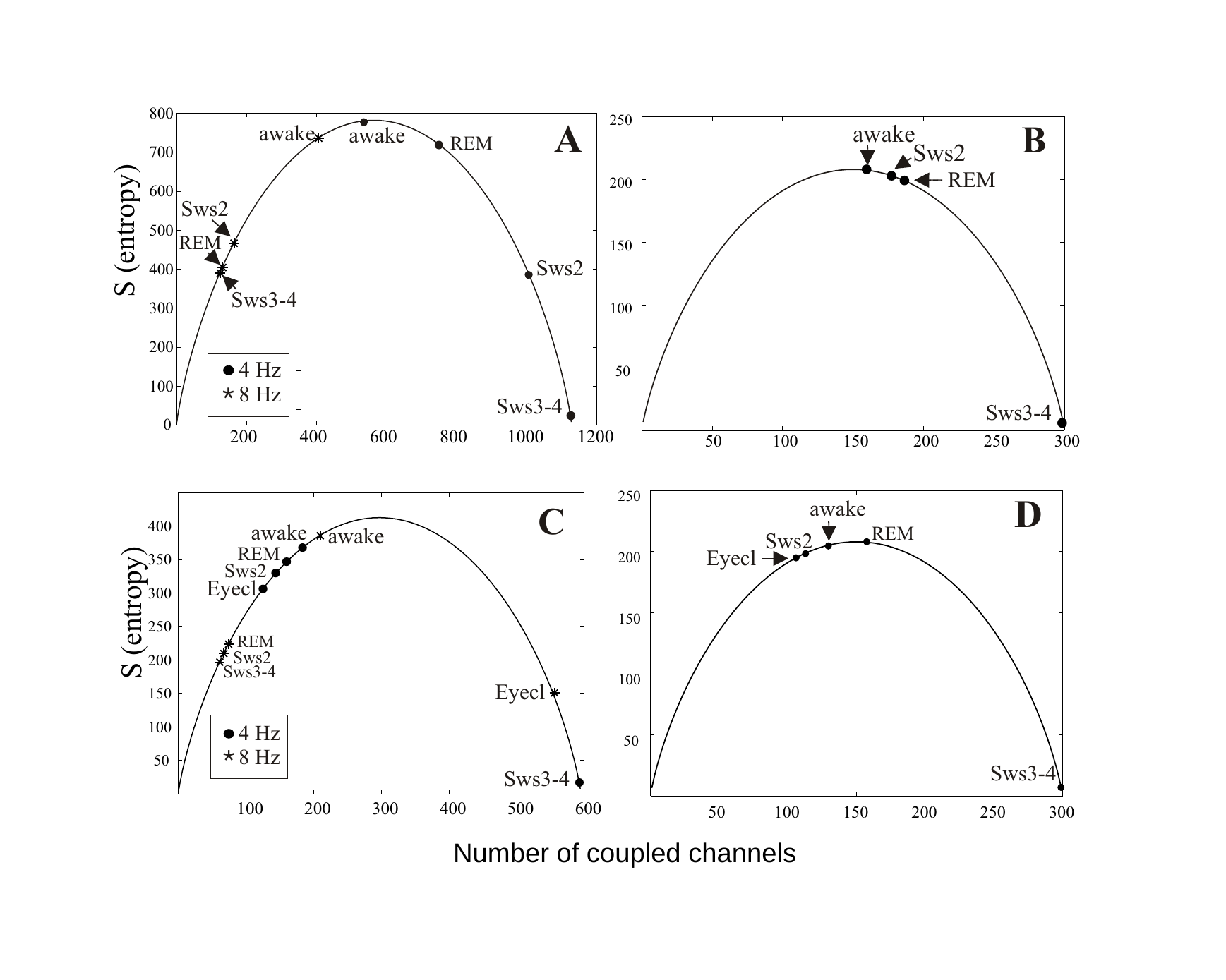}
    
    \caption{ Same graph types as in Figure 1, using sleep recordings. In each subject, data samples were of 2-4 minutes duration during wakefulness with eyes open (‘awake’) or eyes closed (‘Eyecl’), and sleep stages slow-wave 2 (‘Sws2’), slow-wave 3-4 (‘Sws3-4’) and rapid eye movement (‘REM’). A, results derived from iEEG recordings in a subject investigated with bilateral frontal and temporal electrodes and simultaneous scalp EEG. Entropy estimated from synchrony values at two central frequencies of 4 and 8 Hz. As occurred in the patient recordings shown in Figure 1, the baseline, alert state (in this case labelled ‘awake’) is closer to the top of the curve, having greater entropy. The deepest sleep stage, slow wave 3-4 (‘sws3-4’), has the lowest entropy. B, same subject but using the scalp EEG recordings for the calculations, showing similar trend (evaluated at central frequency of 4 Hz). C, results derived from a different subject investigated with right frontal electrodes and simultaneous scalp EEG, with synchrony evaluated at two central frequencies of 4 and 8 Hz using the iEEG signals. Note that the eyes closed (‘Eyecl’) condition has lower entropy than that of the normal alert state with open eyes. Depending on the frequency of analysis, the entropy in ‘Eyecl’ falls toward the left or right side of the curve; e.g., at 8 Hz, because the synchrony is higher (more coupled channels) due to alpha waves at 8-10 Hz and the entropy is reduced due to fewer combinations of connections, as occurs similarly during seizures (Figure 1). D, same subject but using the scalp EEG signals, showing similar results to those obtained with iEEG (evaluated at central frequency of 4 Hz).}  
    \label{figure2}
  \end{center}  
\end{figure}

\begin{figure}[]  
  \begin{center}
    \includegraphics[scale=1.8]{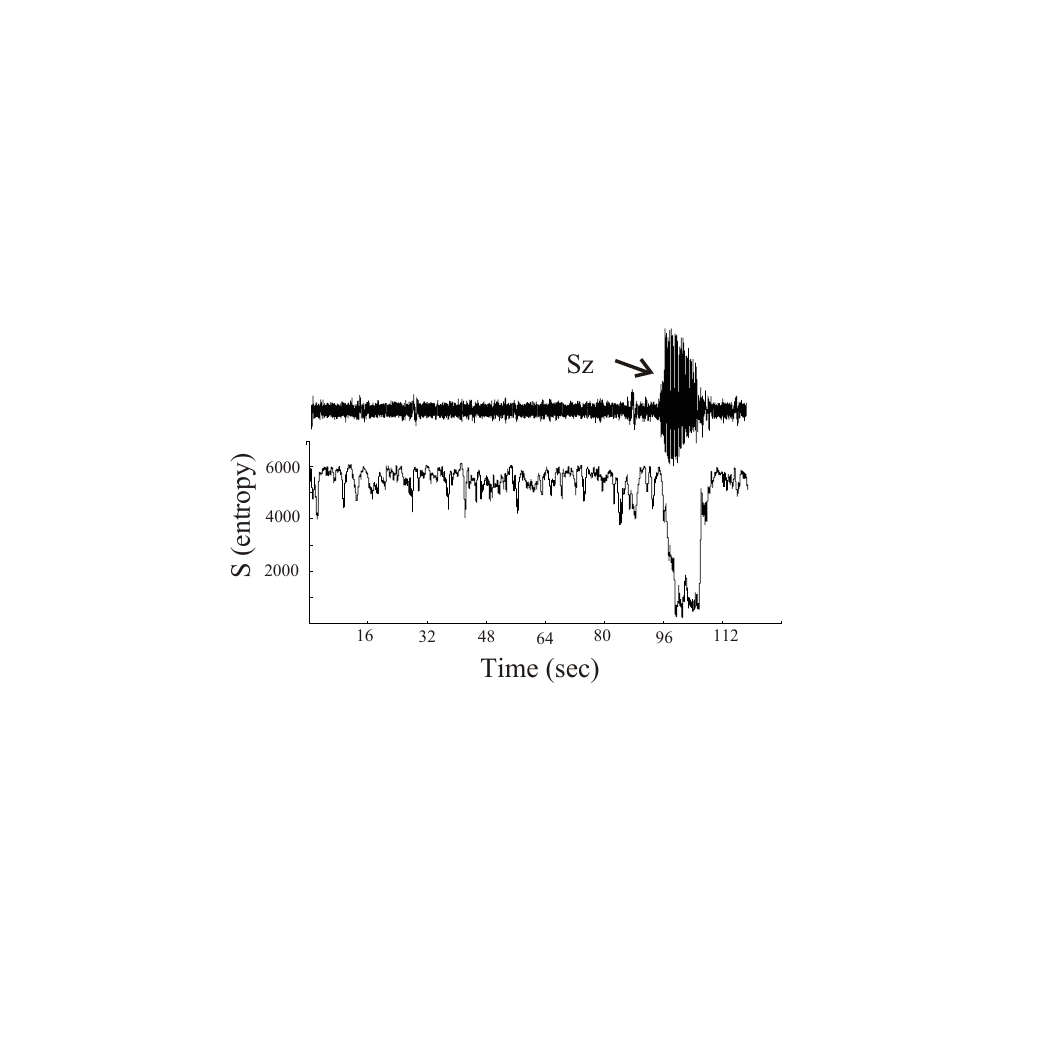}
    
    \caption{ Time course of the entropy of the number of configurations of connected MEG signals before, during and after a generalised absence seizure. MEG signal from one channel is shown at top, the ictus (‘Sz’) occurs towards the end of the ~2 minute recording. Notice the drop in entropy during the seizure.}  
    \label{figure3}
  \end{center}  
\end{figure}

\begin{figure}[]  
  \begin{center}
    \includegraphics[scale=1.5]{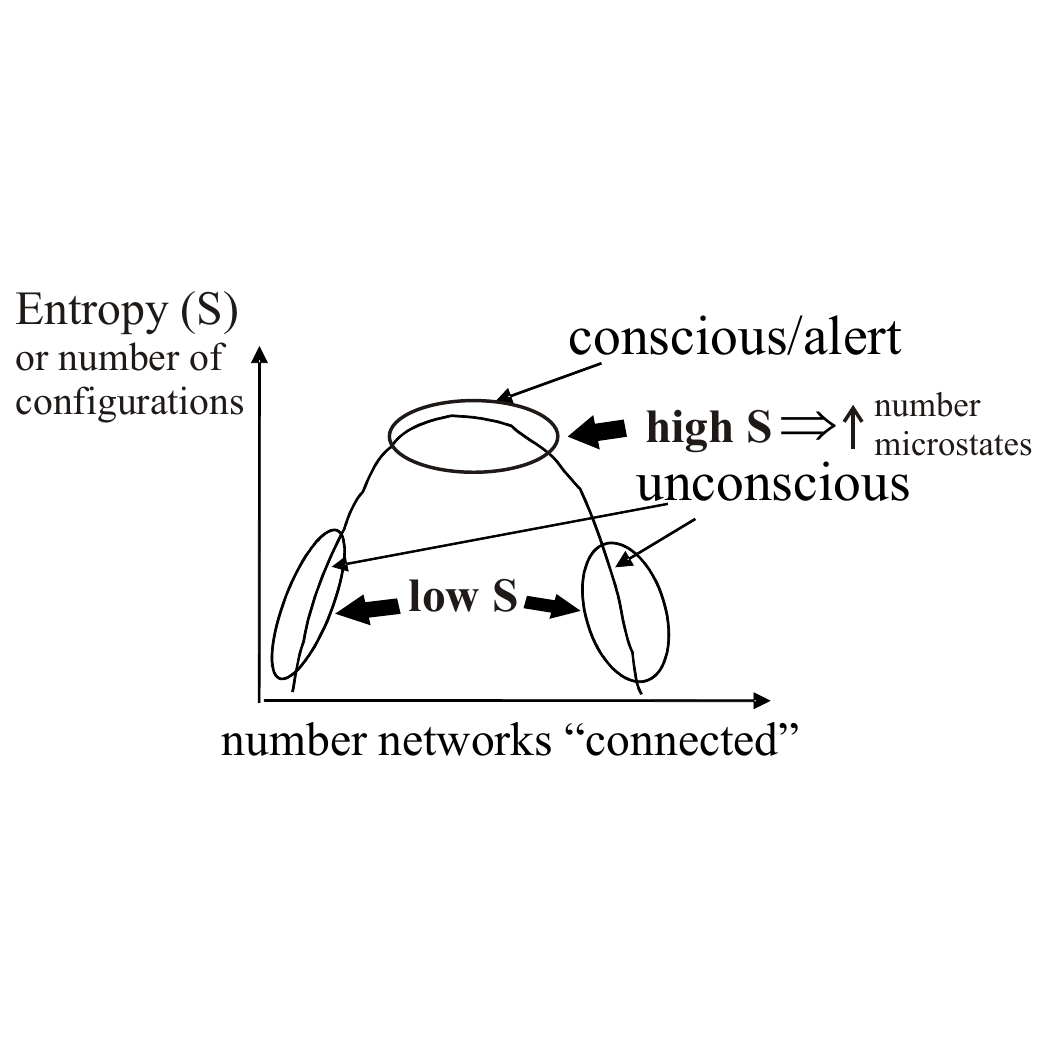}
    
    \caption{ Proposed general scheme of the relation between global brain connectivity and behavioural states. Normal alertness resides at the top of the curve representing the number of configurations of connections the system can adopt, or the associated entropy. The maximisation of the configurations (microstates) provides the variability in brain activity needed for normal sensorimotor action. Abnormal, or unconscious states, are located farther from the top, and are characterised by either large or small number of “connected” networks therefore exhibiting lower number of microstates (hence lower entropy) that are not optimal for sensorimotor processing.}  
    \label{figure4}
  \end{center}  
\end{figure}

  We observe similar trends in the case of sleep. Figure \ref{figure2} depicts some examples. Note how during wakefulness the entropy is closer to the maximum of the curve, whereas the deeper the sleep stage, the more distant from the maximum the values are. The deepest sleep stage, slow wave 3-4 (‘sws3-4’), has consistently the lowest entropy.  Interestingly, the entropy during REM sleep is very close, in most cases, to the normal, alert state. This result may not be as surprising as it sounds if we consider the mental activity during REM episodes that are normally associated with dreams. It is worth noting too that in recordings taken when the subjects had their eyes closed, the entropy is much lower than during the eyes open condition, and sometimes it is as low as during slow-wave 3-4 sleep. The results for two central frequencies (4 and 8 Hz) are shown in Figure \ref{figure2}A and 2C, to demonstrate that it is not always the case of high synchrony having lower entropy; sometimes it is lower synchrony (e.g., results at 8 Hz) that resulted in fewer channel combinations and thus lower entropy as compared to fully alert states. In Figures \ref{figure1} and \ref{figure2} we have shown results with iEEG and MEG (Figure \ref{figure1}) or iEEG and scalp EEG (Figure 2) recordings to demonstrate that the same qualitative results are obtained with these three recordings techniques. Thus these results do not depend on recording methodology. Shown in Figure \ref{figure3} is an example of the time course of the entropy before and during an ictal event. Hence, we have demonstrated that the specific values of the synchrony index R in fully alert states represent the largest number of combinations (microstates, see Discussion) of pairwise signal configurations. This method solves the potential problem of the interpretation of the different R values obtained with various recordings techniques. For example, in our experiments, average values in baseline conditions are $0.248 \pm 0.2$ for MEG, $0.428 \pm 0.04 $for iEEG, and $0.46 \pm 0.05$ for scalp EEG, nevertheless, in our study the number of combinations are “normalised” to the number of recording sensors and show a final result (the entropy) that is independent of the structure and synchrony magnitudes of the recording methodology. The main idea derived from these results is represented in Fig. \ref{figure4}.
  
To further explore whether the decrease in entropy has a parallel with a decrease in other forms of complexity, the Lempel-Ziv complexity of the number of configurations was assessed. Tables 1 and 2 illustrate the Lempel-Ziv complexity estimated for the $B$ matrices, where a complementary result can be seen: unconscious states have lower values of complexity. A decrease of complexity in the raw electrophysiological signals was obtained too (data not shown), hence this may be a phenomenon observable at different levels of description.

\section{Discussion}
Our attempts at seeking features of brain organization that allow for adequate processing of sensory stimuli have provided evidence that a greater number of possible configurations of interactions between brain networks is associated with alert states, representing high entropy associated with the number of those combinations, whereas lower entropy (and thus fewer combinations of connections) is characteristic of either unconscious states or altered states of alertness (eyes closed). This observation reflects a relatively simple general organising principle at this collective level of description, which results in the emergence of properties associated with consciousness.

With the advent of the ‘Big Data’ era and the related torrent of empirical observations, the search for organising principles that result in the emergence of biological phenomena seems more crucial than ever. We tried to uncover relatively simple laws that capture the bounds in the global organization of a biological system that enable it to become adaptable (i.e., responsive) to an environment, or, in neuroscientific terms, the features of optimal brain organization (in terms of connections) that allow brains to adequately process sensory stimuli. We focused on the global states and did not investigate specific patterns of connectivity between brain areas as a variety of other studies have assessed these inter-regional interactions in conscious and unconscious states \cite{dumermuth1981eeg, wang2014eeg}. The fact that our results are similar, independent of recording methodology, demonstrates the robustness of the analysis. On that note, we remark that while we have used the term ‘connectivity’, in reality the analysis reveals only correlation between phases of oscillation, as already discussed in Results.

The present study can be considered an extension of previous work where it was proposed that a general organising principle of natural phenomena is the tendency toward a maximal, or more probable, distribution of energy \cite{velazquez2009finding}, which can be encapsulated by the notion of the maximization of information transfer \cite{haken2006information}. As well, the notions of information and energy exchange are conceptually related: “the common currency paying for all biological information is energy flow” \cite{morowitz1968energy, smith2008thermodynamics}. In the final analysis, information exchange implies energy exchange, hence we interpret information exchange as energy redistribution as proposed in \cite{velazquez2009finding}, even though our study is focused not on energy considerations but on the number of states. Other studies have assessed the importance of brain synchronization to optimally transfer information \cite{buehlmann2010optimal}. We interpret our observation that the number of pairwise channel combinations – that we take as interactions/connections between brain networks – occurs near the maximum of possible configurations in periods with normal alertness, as that greater number of configurations of interactions represents the most probable distribution of energy/information resulting in conscious awareness.  The configuration entropy we calculate measures the information content of the functional network, and has been used in other works for the purpose of quantifying information \cite{bialek1997spikes}. One somewhat surprising result is the low entropy during the eyes-close condition. One could argue that having the eyes closed does not change much the conscious state, and yet we observe reduced entropy as compared with eyes-open condition. Hence our observations may indicate not only states of consciousness but also optimality of sensory processing – considering the great importance of visual processing in humans, interrupting visual inputs should result in considerable changes in the dynamics; for instance, an apparent alteration of brain dynamics upon eye closure is the appearance of alpha waves most clearly in parieto-occipital regions. Interesting too is the relatively high entropy associated with REM episodes, perhaps a reflection of the partial awareness during dreams. Perhaps the main difference between dream and awake states is psychological, but they share similar brain dynamics. 

It has been proposed that aspects of awareness emerge when certain levels of complexity are reached \cite{gell2001consciousness}. It is then possible that the organization (complexity) needed for consciousness to arise requires the maximum number of configurations that allow for a greater variety of interactions between cell assemblies because this structure leads to optimal segregation and integration of information, two fundamental aspects of brain information processing \cite{tononi1998complexity} . Our results help cast the study of consciousness and cognition into more of a physics framework that may provide insight into simple principles guiding the emergence of conscious awareness, and perhaps the proposed thermodynamics for a network of connected neurons \cite{tkacik2014thermodynamics} can be extended to explain cognition. Some classical studies \cite{morowitz1955some} have already characterized biological order in terms of functions of the number of states of a system. It is tempting to link our observations with the typical chemical equilibrium that, despite being composed of a myriad of microstates, when viewed at the macroscopic level produces some useful laws, like the law of mass action describing chemical balance, and thus the perspective we develop here may help guide research to uncover organising principles in the neurosciences. In physics, microstates that yield the same macrostate form an ensemble. A system tends to approach the most probable state, maximising entropy under present constraints, and the resulting macrostate will be represented by the maximum number of microstates. Hence, the macrostate with higher entropy (see scheme in Fig. 4) we have defined, composed of many microstates (the possible combinations of connections between diverse networks, our $C$ variable defined in Results), can be thought of as an ensemble characterised by the largest number of configurations. Here we define an ensemble of microstates as all possible configurations of connectivity leading to the same macrostate (having the same number of connected pair of signals, $p$). The entropy of this macrostate, given by the logarithm of the number of combinations (our $C$), is the number of microstates that are compatible with the given macrostate (assuming all microstates have same statistical weight). In neurophysiological terms, each microstate represents a different connectivity pattern and thus is associated with, in principle, different behaviours or cognitive processes. The macrostate that we find associated with wakeful normal states (eyes open) is the most probable because it has the largest entropy (largest number of combinations of connections). While many elementary cellular microscopic processes are far from equilibrium (e.g., ionic gradients), at the macroscopic level the system tends towards equilibrium, as most natural phenomena remain in near-equilibrium conditions \cite{prigogine1967introduction}. At the same time, the ensemble of microstates associated with normal sensory processing features the most varied configurations and therefore offers the variability needed to optimally process sensory inputs. In this sense, our results support current views on the metastability of brain states \cite{kelso1997dynamic}  in that the states should not be too stable for efficient information processing, hence the larger the number of possible interactions, the more variability is possible. 

Equally, the results are consistent with the global workspace theory \cite{baars1993cognitive}  in that the most widespread distribution of information leads to conscious awareness. Furthermore, these observations relate as well to the information integrated theory \cite{tononi2004information}, in that consciousness increases in proportion to the system’s repertoire of states, thus the more combinations possible, the more states available, and here we can define states as configurations of interactions.  Additionally, the results support computational/theoretical studies showing that patterns of organised activity arise from the maximization of fluctuations in synchrony \cite{vuksanovic2015dynamic}, and that transitions between conscious states are achieved by just varying the probability of connections in neural nets \cite{zhou2015percolation}. In general, our observations highlight the fundamental importance of fluctuations in neuronal activity as the source of healthy brain dynamics. More specifically, our results have a precise parallel with the work of Hudetz et al. (2014), where they graph a dispersion index versus an activation level (their figure 7B) and propose that consciousness resides at the top of the curve, and anaesthetic states and seizures to lower and higher activation levels respectively, as we show in Figures \ref{figure1} and \ref{figure2}. Their ‘activation level’ could correspond to the number of signals that take part in the combinations (our x axes in the graphs), and their ‘dispersion index’ to our number of combinations (the y axes). Other studies have proposed as well that consciousness requires medium values of certain features of cell assemblies \cite{tononi2004information, fingelkurts2014we}. 

In sum, along with others, we consider cognition/consciousness not a static property but a dynamic process with constant flux of energy, or information exchange \cite{haken2006information}. Even though we have talked above about a macrostate, this should not be taken as a fixed state, rather it contains dynamic processes represented by the microstates, the re-arrangements of connections among brain cell ensembles. The emergent features of cognitive phenomena that can be termed “conscious” arise once an efficient web of connections endowed with certain complexity appears. The fact that values of phase synchrony during fully alert states gave us the largest entropy of the number of pairwise signal combinations explains in part the neurophysiological organization underlying these specific values of the quantification of brain synchrony. Studies at this level of description may help to understand how consciousness arises from organization of matter. In our view, consciousness can be considered as an emergent property of the organization of the (embodied) nervous system, especially a consequence of the most probable distribution that maximizes information content of brain functional networks. 

On a technical note, we remark that in our analysis we had to choose a threshold to define “connectivity”, and therefore a suitable baseline had to be chosen. This procedure is obviously biased in that a signal corresponding to a baseline has to be ascribed as the one providing the threshold, and we chose the signals corresponding to the (psychological) state that is most suited for the purposes of adaptability: fully alert and receiving all sensory inputs (awake with open eyes). By this choice, and according to our methods, that signal is already ascribed as having maximum entropy. We tried another less prejudiced method, using surrogates of the original signals, and then computing the average synchrony index among the surrogate population (10 phase-randomised surrogates per original channel/signal) that is the threshold to define connectivity. It turns out that the value of the magnitude of synchrony of the surrogates is close to the one for the aforementioned baseline chosen, so the results do not vary; nevertheless this new method still assigns the largest entropy to the random signals (surrogates), so there is still the assumption that the average synchrony of the stochastic signals is a good approximation to define connections among brain networks. Our current purpose is to completely avoid using thresholds of synchrony indices, for which we are presently working on a scheme that assigns connectivity in the time domain.    

\section{Conclusions and conjectures on the structure of brain-behaviour-environment }

It is tempting to speculate, based on these results and the conclusions of a previous study \cite{velazquez2009finding} that there could be a universal logic ruling the evolution of natural phenomena — biological and nonbiological— and the nervous system in particular: patterns emerge from a central theme captured by maximising information exchange. Because, in the final analysis, all exchange of information implies exchange of energy, natural phenomena tend towards the most probable distribution of energy, and thus the interactions among system constituents tends to be maximized. 

Because the brain functions to maintain a predictive model of the environment (the reason the brain evolved is to model the environment, after all), then perhaps the brain’s global configuration has to “copy” what is out there: and out there energy distributes in all possible microstates (second principle of thermodynamics). Then to process such variability in nature, the nervous system should have same structure, and the result is the ‘inverted U’ that has appeared in our analysis and has been theoretically proposed in other publications \cite{tononi1998complexity}, the top of the curve representing more possible combinations to handle information/energy exchanges. On the other hand, in the extremes of this curve we find fewer microstates, thus these are not optimal situations to process the many microstates in the environment. The key then is not to reach the maximum number of units interacting (which would be all-to-all connections and thus only one possible microstate), but rather the largest possible number of configurations allowed by the constraints. In a similar fashion, it has been argued that the brain needs to show criticality because natural phenomena possess critical dynamics \cite{chialvo2010emergent}. Then, perchance, consciousness can be considered as an emergent property of the organization of the embodied nervous system submerged in an environment, consequence of the most probable distribution of energy (information exchange) in the brain. In this regard, consciousness (like biochemistry) may represent thus an optimal channel for accessing sources of (free) energy.

\newpage
% ==================== Bibliography ==================== %

\bibliography{Sincrony_Bibliografy.bib}
\bibliographystyle{unsrt}

%======================anex=======================

\newpage

\begin{table}[]
\centering

\label{my-label}
\begin{tabular}{|l|l|l|l|l|}
\hline
                  & State & L-Z complexity  \\ \hline
Subject \#1 & Baseline  & 0.7  \\ \hline
                  &  Seizure   & 0.2  \\ \hline
Subject \#2 & Baseline &  0.61 \\ \hline
                  &   Seizure &  0.39  \\ \hline
Subject \#3 & Baseline &  0.86  \\ \hline
                  &  Seizure  &  0.71 \\ \hline
Subject \#4 & Baseline &  0.61  \\ \hline
                   &  Seizure  &  0.61  \\ \hline
\end{tabular}
\caption{Values of Lempel-Ziv (L-Z) complexity derived from the string of connections (details
in Methods) in conscious (baseline) and unconscious (seizure) states, in four patients. Note lower
complexity during seizures in patients $\#~1-3$; patient $\#~ 4$ (Figure 1D) did not have fully generalised
seizures.}
\end{table}
\begin{table}[]
\centering

\label{table_2}
\begin{tabular}{|l|l|l|l|l|}
\hline
                  & State & iEEG L-Z complexity & scalp L-Z complexity  \\ \hline
Subject \#1 & Alert eyes open  & 0.81& 1.01\\ \hline
                  &  SWS~2  & 0.42 & 1.0\\ \hline
                  & SWS~3-4 &  0.018 & 0.0 \\ \hline
                  &  REM  &  0.73 & 0.96 \\ \hline
Subject \#2 & Alert eyes open  & 0.94 & 1.06 \\ \hline
                  &  Alert eyes close & 0.94 & 0.85 \\ \hline
                  & SWS~3-4 & 0.33 & 0.22 \\ \hline
Subject \#3 & Alert eyes open  & N/A & 1.07 \\ \hline
                  &  SWS~2  & N/A & 1.07\\ \hline
                  & SWS~3-4 & N/A & 1.07\\ \hline
                  &  REM  & N/A & 0.97\\ \hline
Subject \#4 & Alert eyes open  & 0.88 & 1.0\\ \hline
                  & Alert eyes closed  & 0.55 & 1.0\\ \hline
                  &  SWS~2  & 0.57 & 9.8\\ \hline
                  & SWS~3 & 0.05 & 0.0\\ \hline
                  &  REM  & 0.6  & 1.06\\ \hline
Subject \#5 & Alert eyes open  &  N/A & 0.97 \\ \hline
                  &  SWS~1  & N/A & 1.08\\ \hline
                  & SWS~2 & N/A  & 1.04 \\ \hline
                  &  SWS~3-4  & N/A & 0.8\\ \hline
                  &  REM  & N/A & 1.12\\ \hline
\end{tabular}
\caption{Values of Lempel-Ziv (L-Z) complexity derived from the string of connections in
different sleep stages. When both recordings were obtained from a subject, both iEEG and scalp
EEG data were analysed (subjects $\#~3$ and $\#~5$ did not have iEEG recordings). The L-Z complexity
is consistently lower, regardless of recording methodology, in the deepest sleep stage (SWS 3-4).
These results parallel those of entropy estimations shown in figure 2}
\end{table}

%===============================Figures=============================================================================
\newpage

%-------------------------------------------------------------------------------------------------------------------------

%-------------------------------------------------------------------------------------------------------------------------

%-------------------------------------------------------------------------------------------------------------------------

%-------------------------------------------------------------------------------------------------------------------------

\end{document}